\documentclass[amsmath,amssymb,nofootinbib,superscriptaddress]{revtex4-1}

\usepackage{aas_macros,esint,graphicx}

\usepackage{soul}
\usepackage{comment}
\usepackage[compact]{titlesec}         
\titlespacing{\subsection}{10pt}{10pt}{10pt}
\titlespacing{\section}{10pt}{10pt}{10pt}

\usepackage[allcolors=blue,colorlinks=true]{hyperref}
\numberwithin{equation}{section}

\makeatletter
\renewcommand{\p@subsection}{}
\renewcommand{\p@subsubsection}{}
\makeatother

\newcommand{\br}[1]{\left[#1\right]}

\newcommand{\pa}[1]{\left(#1\right)}

\newcommand{\ed}{\mathop{}\!\mathrm{d}}

\begin{document}

\title{\vspace*{40pt}\huge{Logarithmic Corrections to Kerr Thermodynamics}\vspace*{40pt}}

\author{Daniel Kapec}
\email{danielkapec@fas.harvard.edu}
\affiliation{Harvard Center of Mathematical Sciences and Applications, Harvard University, Cambridge, MA 02138}
\affiliation{Center for the Fundamental Laws of Nature, Harvard University, Cambridge, MA 02138, USA}

\author{Ahmed Sheta}
\email{asheta@g.harvard.edu}
\affiliation{Center for the Fundamental Laws of Nature, Harvard University, Cambridge, MA 02138, USA}
\author{Andrew Strominger}
\email{strominger@physics.harvard.edu}
\affiliation{Center for the Fundamental Laws of Nature, Harvard University, Cambridge, MA 02138, USA}
\author{Chiara Toldo}
\email{chiaratoldo@fas.harvard.edu}
\affiliation{Center for the Fundamental Laws of Nature, Harvard University, Cambridge, MA 02138, USA}
\affiliation{Dipartimento di Fisica, Universita' di Milano, via Celoria 6, 20133 Milano MI, Italy
}

\begin{abstract}
\vspace*{40pt}
Recent work has shown that loop corrections from massless particles generate $\frac{3}{2}\log T_{\text{Hawking}}$  corrections to black hole entropy which dominate the thermodynamics of cold near-extreme charged black holes. Here we adapt this analysis to near-extreme Kerr black holes.
Like AdS$_2\times S^2$, the Near-Horizon Extreme Kerr (NHEK) metric has a family of normalizable zero modes corresponding to reparametrizations of boundary time. The path integral over these zero modes leads to an infrared divergence in the one-loop approximation to the Euclidean NHEK partition function. We regulate this divergence by retaining the leading finite temperature correction in the NHEK scaling limit. This ``not-NHEK'' geometry lifts the eigenvalues of the zero modes, rendering the path integral infrared finite. The quantum-corrected near-extremal entropy exhibits $\frac{3}{2}\log T_{\text{Hawking}}$ behavior characteristic of the Schwarzian model and predicts a lifting of the ground state degeneracy for the extremal Kerr black hole. 
\end{abstract}

\maketitle

\clearpage

{
  \hypersetup{linkcolor=black}
  \tableofcontents
}

\vspace{30pt}

\section{Introduction}
To an outside observer, a black hole appears to be an ordinary quantum mechanical system with finite entropy and highly chaotic internal dynamics. 
According to this picture, the exponential of the Bekenstein-Hawking entropy $e^{S_{\text{BH}}}$ represents the smooth (coarse-grained) leading approximation to the density of states of the black hole Hilbert space, whose average level spacing is expected to be $e^{-S_{\text{BH}}}$. 
The  near-equilibrium dynamics  of a black hole in a pure microstate is believed to be well-approximated by statistical averages in the canonical  or microcanonical ensemble since the Hamiltonian is chaotic. In the special cases for which an explicit microscopic description is available, these averages are computed using standard statistical techniques and are found to agree with the predictions of semi-classical general relativity and Euclidean quantum gravity. When no explicit microscopic description is available (as is most often the case), the averages can only be performed macroscopically by evaluating the Euclidean gravitational path integral in a saddle point approximation. 

Although black holes are believed to be ``ordinary'' quantum mechanical systems, their thermodynamics is not generic. Black holes that spin or carry charge can be very large and very cold, and in the leading order semiclassical approximation to the black hole density of states there is  an enormous ground state degeneracy $e^{S_0}$ for these systems. 
In theories with unbroken supersymmetry, the existence of these ground states is sensible due to the huge degeneracy at zero coupling where there is enhanced symmetry, but in less symmetric models (like the black holes in our universe!) the degeneracy is surprising and one wants to know if it is merely the consequence of an approximation or not. This paper addresses this question for Kerr black holes of spin $J$, which are extremal when $J=M^2$ with entropy $2\pi J$.

This question is related to another old puzzle about cold back holes \cite{Preskill:1991tb}.
For a black hole near extremality, semiclassical analysis predicts that the thermodynamically accessible energy above extremality $E_{\text{BH}}$ scales quadratically with the temperature of the black hole,  while the typical energy $E_{\text{Hawk}}$ of a quantum  of Hawking radiation scales linearly
\begin{equation}\label{eq:puzzle}
    E_{\text{BH}}(T) \sim J^{3/2}T^2 \; , \qquad     E_{\text{Hawk}} \sim T \; .
\end{equation}
Below the temperature where these two curves intersect, the emission of a thermal Hawking quantum  cannot be treated as a near-equilibrium process, since the emission of such a quantum would carry away all of the energy available in the system. In other words, even though the curvature at the horizon is small, the semiclassical analysis does not seem to apply. This puzzle was resolved in \cite{Preskill:1991tb} by showing that the leading semiclassical approximation receives large corrections below $E_{\text{BH}}=E_{\text{Hawk}}$ and therefore cannot be trusted, but a derivation  of the behavior in this regime was not given.

There are now two proposed behaviors, both of which appear to have realizations in different models.  The authors of \cite{Preskill:1991tb} suggested that the black hole spectrum might have a gap $E_{\text{gap}}\sim J^{-3/2}$ above extremality, below which thermodynamics obviously no longer applies. For black holes with known microscopic descriptions (all of which are supersymmetric) this gap indeed exists \cite{Maldacena:1997ih,Iliesiu:2020qvm,Heydeman:2020hhw,Boruch:2022tno,Iliesiu:2022kny,Iliesiu:2022onk,Sen:2023dps}. The quantum mechanics that describes these black holes is continuously connected to a free point with enhanced symmetry and many BPS states which are protected all the way to strong coupling, so it is not surprising that these models have a large ground state degeneracy in a fixed charge sector. 
In the absence of supersymmetry it is less clear what to expect.

The second possibility is that the large ground state degeneracy is an artifact of the leading order calculation, and that quantum corrections become more relevant at low temperatures and cause these states to spread out over a dense energy band above the vacuum. This is what one would naively expect for a non-supersymmetric system like the Kerr black hole, the focus of this paper. Although the exact spectrum of the black hole can only be computed non-perturbatively (the expected eigenvalue spacings in this part of the spectrum are $e^{-S_{\text{BH}}}\sim e^{-1/G_N}$), in this scenario one hopes to compute a perturbative correction to the density of states and determine whether or not $\rho_{\text{corr}}(E)\to 0$ as $E\to 0$. 

Calculating this correction directly using the full Kerr geometry is a formidable task so far unachieved for any near-extremal black hole. However, there is another approach for studying the low temperature thermodynamics of spinning black holes that makes use of the emergent near-extremal throat (NHEK) and its approximate decoupling from the far region. At exact extremality the throat is infinitely long, and its asymptotic boundary serves as an effective stretched horizon for the black hole system. In this limit the far region  decouples\footnote{Obstacles to decoupling, not considered  here,  may arise from superradiant modes.}  and one expects that the relevant part of the black hole Hilbert space can be equivalently captured by gravitational dynamics in the throat according to the Kerr/CFT correspondence. The analogous formalism, when applied to spherically symmetric black holes, has led to precise matches of bulk gravitational calculations and microscopic counts  \cite{Sen:2014aja}. 

However, using this approach to study the excited near-extremal microstates is subtle. In particular, due to the strong backreaction effects present in low dimensional systems with long-range forces, quantum gravity with exactly AdS$_2$ boundary conditions is believed to only describe ground states \cite{Maldacena:1998uz,Amsel:2009ev}. Calculations involving excited states are beset with infrared (IR) divergences, indicating a failure of the black hole to fully decouple from the far region. 

In fact, as first noted by Sen \cite{Sen:2012cj}, even the ground-state calculations can suffer from subtle divergences.  The simplest IR divergence manifests in the one-loop correction to the Euclidean partition function in the extremal throat. In the process of calculating logarithmic corrections to extremal black hole entropy, Sen identified a set of normalizable zero modes in the NHEK throat corresponding to $\text{Diff}(S^1)/SL(2,\mathbb{R})$ diffeomorphisms with non-compact support. Since these fluctuations are normalizable they must be integrated over, and since the domain of integration is infinite dimensional with no suppression the partition function diverges
\begin{equation}\label{eq:ZMpi}
    Z_{\text{NHEK}}\propto \int \displaylimits_{\text{Diff}(S^1)/SL(2,\mathbb{R})}[Dh] \;  = \infty \; .
\end{equation}
The dependence of the measure on $S_0$ can be unambiguously determined, so these zero modes contribute a known logarithmic (in $S_0$) correction to the extremal entropy, assuming that it exists.
However, the IR divergence of the partition function due to the unsuppressed fluctuations of the zero modes signals a subtlety in the calculation. As we will see, a proper treatment of these zero  modes can remove the ground state degeneracy entirely, rendering the $T\to 0$ limit of the quantum black hole singular.
Instead of a system with tremendous entropy at zero temperature, one encounters a system with a dense energy band of $e^{S_0}$ states spread out above the vacuum to which standard thermodynamics applies. The corresponding correction to the thermodynamic energy alleviates the tension inherent in \eqref{eq:puzzle}.

This resolution to the puzzle raised by \eqref{eq:ZMpi} was first proposed for the analogous problem in the AdS$_2$ throat of extreme Reissner-Nordstr\"{o}m  in \cite{Iliesiu:2022onk,Banerjee:2023quv}, and our analysis follows theirs closely.  The strategy adopted in these papers amounts to turning on a small but finite temperature $T$, which necessitates the retention of subleading corrections to the metric in the near-extremal throat. These metric corrections lift the zero mode degeneracy and lead to log $T$ corrections to the near-extremal entropy that agree with results derived more indirectly  using the Schwarzian model \cite{Maldacena:2016upp}. Laplace transforming this result to obtain the density of states, one finds that the ground state degeneracy is actually sub-exponential in $S_0$ (it vanishes at this order in the approximation but presumably receives non-perturbative corrections).
It is this prescription that we adapt for the near-extreme Kerr black hole, as described below. 

The standard scaling limit into the throat of the extreme Kerr black hole takes the form
\begin{align}\label{eq:Scalinglimit}
	\hat{t}=\frac{1}{2\pi T}t,\qquad
	\hat{r}=r_+(T)+4\pi r_0^2 T  (r-1),\qquad
        \hat{\phi}=\phi +\frac{t}{4\pi r_0 T}-t	\; ,
         \qquad T\to 0 \; ,
\end{align}
leading to the decoupled NHEK metric
\begin{equation}\label{eq:NHEK}
	ds^2=J(1+\cos ^2\theta)\left(-(r^2-1)\ed t^2+\frac{\ed r^2}{r^2-1}+\ed\theta^2\right)+
 J\frac{4\sin^2\theta}{1+\cos^2\theta}\left(\ed\phi+(r-1)\ed t\right)^2.
\end{equation}
Here $J$ is the spin, $r_+(T)$ is the radius of the outer horizon,  $r_0$ is the radius of the extremal horizon and we take the limit in Boyer-Lindquist coordinates $(\hat{t},\hat{r},\theta,\hat{\phi})$. 
The Euclidean continuation of this metric has zero modes
which lead to the infrared divergence \eqref{eq:ZMpi}. If one retains the leading $O(T)$ correction to this metric in the scaling limit \eqref{eq:Scalinglimit}, one obtains a distinct geometry which we will term the ``not-NHEK'' metric\footnote{The terminology near-NHEK is already commonly used to denote the leading order geometry \eqref{eq:NHEK}, which is really a Rindler patch of the full global NHEK geometry. It should not be confused with the term ``nearly-AdS$_2$'' which describes the Reissner-Nordstr\"{o}m analog of the metric \eqref{eq:IntroNotNHEK}.}
\begin{equation}\label{eq:IntroNotNHEK}
    g_{\text{not-NHEK}}=g_{\text{NHEK}}+  \delta g \; ,
\end{equation}
with $\delta g \sim T$. 
Unlike \eqref{eq:NHEK}, this metric is not an exact solution to the four-dimensional Einstein equation, although one can view it as a perturbative (in $T$) approximation to a solution whose nonlinear completion is the asymptotically flat finite temperature black hole. Indeed, the second term in \eqref{eq:IntroNotNHEK} is interpreted as the leading approximation to the far-region metric as seen by the near-horizon observer, and as such it couples the black hole throat dynamics to the asymptotically flat spacetime. It is easy to see that the zero-modes of \eqref{eq:NHEK} that lead to the divergence of \eqref{eq:ZMpi} are lifted by the perturbation \eqref{eq:IntroNotNHEK}. The normalizable zero modes identified by Sen  are metric deformations generated by non-normalizable diffeomorphisms with non-compact support, meaning that they can be written 
\begin{equation}
    h^{(n)}=\mathcal{L}_{\xi^{(n)}}g_{\text{NHEK}}
\end{equation}
for non-normalizable vector fields $\xi^{(n)}$. These vector fields  are closely related to  Virasoro generators of Kerr/CFT \cite{Guica:2008mu,Castro:2009jf,Bredberg:2011hp}\footnote{Similarly the ones appearing in the Reissner-Nordstr\"{o}m case are related to those in RN/CFT \cite{Hartman:2008pb}.}, possibly as twisted in \cite{Hartman:2008dq} (a connection we hope to explore further). However, they are \textit{not} diffeomorphisms of the not-NHEK metric
\begin{equation}\label{eq:zmnotNHEK}
    h^{(n)}\neq\mathcal{L}_{\zeta}g_{\text{not-NHEK}} \; .
\end{equation}
They therefore acquire temperature-dependent eigenvalues at first order in perturbation theory. These perturbed eigenvalues can be used to obtain an approximation for the small-$T$ (zeta-regularized) Euclidean partition function in the not-NHEK geometry. Interpreted as a correction to the black hole partition function, these new terms predict a lifting of the extremal ground state degeneracy for the Kerr black hole and a resolution of the puzzle described in \cite{Preskill:1991tb}.

There are many  subtleties both in the calculation of logarithmic corrections to (near)-extreme black hole thermodynamics and in the physical interpretation of the results. There are both gauge and geometric ambiguities in the ``gluing" of the decoupled near horizon geometry to the asymptotically flat region. The superradiant instability of the NHEK throat leads to  travelling waves with imaginary conformal weights and a complex partition function, and calls into question the exact decoupling of the two regions, as does the nonexistence of a global vacuum for Kerr.  Black holes in asymptotically flat spacetimes have finite lifetimes and are therefore metastable resonances rather than eigenstates. Since the ``eigenvalue'' spacing for these black holes is roughly $ e^{-S_0}$ while the lifetime is polynomial in $S_0$, the widths are naively much larger than the spacings and it is not clear whether it is sensible to discuss a discrete density of states. 
Another  question regards the relationship between \eqref{eq:IntroNotNHEK} and the lifting of the zero modes in the zero temperature throat. One could equally well keep the leading order correction to the NHEK metric in the scaling limit of an exactly extremal  black hole \cite{Hadar:2020kry}.
This potentially  allows one to lift the zero modes and define a finite zero temperature partition function associated to the black hole, in contradiction with the statement that the $T\to 0$ limit is always singular.  There is also the choice of ensemble. While leading order semiclassical results  are insensitive to these subtleties, the situation for  subleading effects is still under discussion.

While not all of these issues have been definitively settled, it is nevertheless clear that significant recent progress has been made in the understanding of logarithmic corrections to near-extreme charged black hole thermodynamics and extreme Kerr thermodynamics.  The purpose of this paper is to fill in the missing analysis of $near$-extreme Kerr thermodynamics by simply  adopting both the assumptions and methodology used for the Reissner-Nordstr\"{o}m case in the seminal papers \cite{Iliesiu:2022onk,Banerjee:2023quv}.  Although the details differ, at the end we interestingly find a numerically identical entropy shift  of ${3 \over 2}\log T$, compatible with Schwarzian dynamics. Our main mathematical results are formulas for the finite temperature eigenvalues of the NHEK zero modes in the not-NHEK geometry \eqref{corr_e}, whose detailed form leads to the  factor $\frac{3}{2}$ in \eqref{corr_Z}.

The outline of this paper is as follows. Section \ref{sec:KerrReview} reviews Kerr thermodynamics, focusing on the small temperature expansion and near-horizon limit. Section \ref{sec:qef} introduces the corrected ``not-NHEK'' geometry and calculates the correction to the NHEK partition function. Appendix \ref{app:AdS2} reviews the analogous AdS$_2$ calculations.

As this paper was nearing completion, we became aware that similar results are in preparation by I. Rakic, M. Rangamani and G. J. Turiaci \cite{Rakic}.

\section{Near-Extreme Kerr}\label{sec:KerrReview}

The Kerr metric in Boyer-Lindquist coordinates takes the form
\begin{subequations}
\label{eq:Kerr}
\begin{gather}
	ds^2=-\frac{\Delta}{\Sigma}\pa{\ed \hat{t}-a\sin^2{\theta}\ed\hat{\phi}}^2+\frac{\Sigma}{\Delta}\ed \hat{r}^2+\Sigma\ed\theta^2+\frac{\sin^2{\theta}}{\Sigma}\br{\pa{\hat{r}^2+a^2}\ed\hat{\phi}-a\ed \hat{t}}^2,\\
	\Delta(\hat{r})=\hat{r}^2-2M\hat{r}+a^2,\qquad
	\Sigma(\hat{r},\theta)=\hat{r}^2+a^2\cos^2{\theta}. \notag
\end{gather}
\end{subequations}
The spin of the black hole is given by $J=aM$, the inner and outer horizons occur at $r_{\pm}= M  \pm \sqrt{M^2-a^2}$,
and the area of the outer event horizon is
\begin{equation}
    A=4\pi(r_+^2+a^2)=8\pi M \left( M+ \sqrt{M^2-(J/M)^2}\right) \; .
\end{equation}
The Hawking temperature and angular velocity of the horizon are
\begin{equation}\label{eq:kerrT}
  T  =\frac{1}{4 \pi M}\frac{\sqrt{M^2-(J/M)^2}}{M+\sqrt{M^2-(J/M)^2}} \; , \qquad  \Omega_H=\frac{a}{2Mr_+} \; .
\end{equation}
In the extremal limit $M^2\to M_0^2=J$, the horizons coalesce at $r_0=M_0$, the temperature vanishes, and $\Omega_H\to \frac{1}{2r_0}$. 

At fixed angular momentum $J=r_0^2$, we parameterize small deviations from extremality by their temperature $T$. The relation (\ref{eq:kerrT}) defines the thermodynamic energy $M(J,T)$, which has the small temperature expansion
\begin{equation}
    M(T,J)=J^{1/2} + 4\pi^2 J^{3/2}T^2+32\pi^3 J^2T^3 + 264\pi^4 J^{5/2}T^4 + \dots
\end{equation}
Similarly, the horizons $r_{\pm}(T)$ at fixed $T$ have the small temperature expansion
\begin{equation}
    \begin{aligned}
        &r_+(T) = J^{1/2} + 4\pi JT + 20 \pi^2 J^{3/2}T^2 +128\pi^3 J^2 T^3 + 968\pi^4 J^{5/2}T^4+ \dots \; , \\
        &r_-(T) =J^{1/2}-4 \pi  J T-12 \pi ^2 J^{3/2} T^2-64 \pi ^3 J^2 T^3 -440 \pi ^4 J^{5/2} T^4 +\dots
    \end{aligned}
\end{equation}
The near-extremal entropy is then linear in $T$
\begin{equation}\label{eq:LinT}
    S(T,J)= S_0 + 8\pi^2J^{3/2}T  + O(T^2)  \; ,
\end{equation}
and the average thermodynamic energy above extremality scales quadratically with temperature as 
\begin{equation}
    E(T,J) = M(T,J) - M_0 = 4\pi^2 J^{3/2}T^2  + O(T^3) \; .
\end{equation}

\subsection{The NHEK Throat}
For many purposes one would like to study the dynamics of the black hole as an isolated quantum system, independent of its embedding in the full asymptotically flat spacetime. Unfortunately, for a generic Kerr black hole, there is no meaningful geometric separation between the region of spacetime associated to the hole and the spacetime belonging to the far region: the two systems are coupled and the interactions between them cannot be ignored. The exception occurs when the black hole is near-extremal, in which case a long throat of length $|\log T|$ develops just outside of the horizon.  In the limit of infinite proper depth, this region is believed to approximately decouple from the far region, although  this itself is a subtle statement. This region of spacetime is generally associated to the black hole. 

In practice, it is possible to isolate the extremal throat by taking a scaling limit that zooms into the near horizon region of a family of cold Kerr geometries. The change of coordinates
\begin{equation} \label{eq:throat diff}
\begin{gathered}
    \hat{t}=\frac{2r_0}{\varepsilon(T)}t,\qquad
	\hat{r}=r_+(T)+r_0 \varepsilon(T) (\cosh \eta-1),\qquad
        \hat{\phi}=\phi +\frac{t}{\varepsilon(T)}-t	\; , \qquad  \varepsilon(T) = 4 \pi r_0 T,
\end{gathered}
\end{equation}
followed by the limit $T\to0$, results in a spacetime that solves the Einstein equation in its own right:
\begin{equation} \label{eq:NHEK_sen}
     ds^2=J \left(1+\cos ^2\theta \right) \left(-\sinh ^2\eta \,dt^2+d\eta^2+d\theta ^2  \right)+ 
     J \frac{4 \sin ^2\theta  }{1+\cos ^2\theta }(d\phi+(\cosh \eta-1)\, dt   )^2 \; .
\end{equation}
This geometry, found by Bardeen and Horowitz in \cite{Bardeen:1999px}, is known as Near-Horizon Extreme Kerr (NHEK).\footnote{Some references refer to this spacetime as near-NHEK, and reserve the term NHEK for its geodesic completion. } It  is the analog of the Robinson-Bertotti universe obtained from the scaling limit of the near-extremal Reissner-Nordstr\"{o}m black hole. The metric has $SL(2,\mathbb{R})\times U(1)$ symmetry with generators 
\begin{equation}\label{eq:NHEKSl2}
    L_{\pm 1}= \frac{e^{\mp t}}{\sinh \eta}(\cosh \eta \, \partial_t \pm \sinh \eta \,\partial_\eta +(\cosh\eta -1)\partial_{\phi})\;, \qquad L_0= \partial_t+\partial_\phi\;, \qquad W=\partial_\phi.
\end{equation}
It is commonly believed that at least part of the quantum mechanics of the Kerr black hole is captured by gravitational dynamics in this throat in analogy with the better-understood black holes with near-horizon AdS regions.

\section{Quantum Corrections to the Throat Thermodynamics}
\label{sec:qef}
There is to date no top-down microscopic construction of the four-dimensional Kerr black hole.\footnote{Although embeddings of nonsupersymmetric Kerr-like rotating black holes in string theory can be found in \cite{Guica:2010ej,Compere:2010uk}.}  However in accord with the usual assumptions we will  identify the  analytically continued gravitational path integral  in the NHEK throat with the statistical partition function of the dual quantum mechanics.

Following Sen \cite{Sen:2012cj} we analytically continue  $t=-i\tau$ in (\ref{eq:NHEK_sen}) which gives \begin{equation}\label{eq:EucNHEK}
    ds^2=J(1+\cos^2\theta)(d\eta^2 +\sinh^2 \eta d\tau^2 + d\theta^2) + J \frac{4\sin^2\theta}{1+\cos^2\theta}(d\phi -i(\cosh \eta -1)d\tau)^2 \;.
\end{equation}
Regularity of the geometry at $\eta = 0$ requires the periodicity $\tau \sim \tau+2\pi$. 
The partition function in the near-horizon region of Kerr is given formally by an integral over metrics subject to a certain set of boundary conditions \cite{Sen:2012cj}
\begin{equation} \label{eq:ZNHEK}
    Z = \int [Dg] e^{-I[g]} \; \qquad I[g]=-\frac{1}{16 \pi} \int_{\mathcal{M}} d^4x \sqrt{g} R + I_{\text{boundary}} \; .
\end{equation}
The boundary conditions and corresponding boundary terms in the action determine the statistical ensemble computed by the path integral.
 
The geometry \eqref{eq:EucNHEK} is a classical saddle-point for the integral \eqref{eq:ZNHEK} satisfying the appropriate boundary conditions, and therefore provides the leading approximation $Z\approx \exp\left(-I[g_{NHEK}]\right)$ to the black hole partition function. In \cite{Sen:2012cj} it was shown that this saddle-point approximation, including the correct boundary contributions, reproduces the semiclassical entropy $S_0$ of the extremal Kerr black hole. However, the path integral \eqref{eq:ZNHEK} is not well-defined beyond the leading saddle point approximation: it is beset with UV divergences, and the instability of the NHEK throat due to superradiance means that any sensible definition of the integral will necessarily make $Z$ complex. 
Nevertheless, in \cite{Sen:2012cj} Sen managed to extract some universal information about the dependence of \eqref{eq:ZNHEK} on $S_0$ through a careful analysis of the 1-loop determinant of massless fields on the background \eqref{eq:EucNHEK}. These logarithmic corrections to the Kerr black hole entropy   provide a stringent test for any proposed microscopic dual to the Kerr black hole. 

\subsection{Quantum Corrections to NHEK Entropy and Zero Modes}
The determination of the logarithmic corrections to the extremal Kerr entropy requires path-integration over the massless fields propagating on the NHEK throat. Expanding about the saddle-point $g=\bar{g}+h$, with $\bar{g}$ given by \eqref{eq:EucNHEK} and $h$ a normalizable perturbation, the 1-loop approximation is controlled by the linearized kinetic operator for $h$
\begin{equation}\label{eq:1loop}
   Z\approx \exp\left(-I[\bar{g}]\right)\int [Dh]\exp \left[-\int d^4x \sqrt{\bar{g}}\,hD[\bar{g}]h\right] \; .
\end{equation}
Calculations are performed with the gauge fixing term
\begin{equation}\label{eq:GF}
     \mathcal{L}_{GF} = \frac{1}{32 \pi} \bar{g}_{\mu \nu} \left( \bar{\nabla}_\alpha h^{\alpha \mu} - \frac12 \bar{\nabla}^\mu h^{\alpha \;} _{\;\alpha} \right) \left( \bar{\nabla}_\beta h^{\beta \nu} - \frac12 \bar{\nabla}^\nu h^{\beta \;} _{\; \beta} \right) \; 
\end{equation}
which, when combined with the Einstein-Hilbert action, yields the linearized kinetic term \cite{Bhattacharyya:2012wz}
\begin{equation}\label{eq:gravKinOp}
    h_{\alpha \beta} D^{\alpha \beta, \mu \nu} [\bar{g}] h_{\mu \nu} = -\frac{1}{16 \pi} h_{\alpha \beta} \left(\frac14 \bar{g}^{\alpha \mu} \bar{g}^{\beta \nu} \bar{\Box} -\frac18 \bar{g}^{\alpha \beta} \bar{g}^{\mu \nu} \bar{\Box} +\frac12 \bar{R}^{\alpha \mu \beta \nu} +\frac12 \bar{R}^{\alpha \mu} \bar{g}^{\beta \nu} - \frac12 \bar{R}^{\alpha \beta} \bar{g}^{\mu \nu}  -\frac14 \bar{R} \bar{g}^{\alpha \mu} \bar{g}^{\beta \nu} + \frac18 \bar{R} \bar{g}^{\alpha \beta} \bar{g}^{\mu \nu}   \right) h_{\mu \nu} \; .
\end{equation}
For NHEK this is simply 
\begin{equation}\label{eq:NhekKinOp}
    h_{\alpha \beta} D_{\text{NHEK}}^{\alpha \beta, \mu \nu}  h_{\mu \nu} = -\frac{1}{16 \pi} h_{\alpha \beta} \left(\frac14 \bar{g}^{\alpha \mu} \bar{g}^{\beta \nu} \bar{\Box} -\frac18 \bar{g}^{\alpha \beta} \bar{g}^{\mu \nu} \bar{\Box} +\frac12 \bar{R}^{\alpha \mu \beta \nu} \right) h_{\mu \nu} \;  .
\end{equation}
The determinant of this operator cannot be calculated exactly due to the reduced symmetry of the problem, but the terms contributing logarithmic corrections in $S_0$ can be extracted indirectly through the heat kernel expansion and the appropriate integrals over curvature invariants. The details of this calculation are not directly relevant for what follows. What is important for the present discussion is the fact that the operator appearing in \eqref{eq:NhekKinOp} supports a family of normalizable zero modes
\begin{equation}\label{eq:ZmodesKERR}
    h_{\mu \nu}^{(n)}dx^\mu dx^\nu= \frac{1}{4 \pi} \sqrt{\frac{3}{2}} \sqrt{|n|(n^2-1)}(1 + \cos^2\theta)e^{in \tau} \frac{(\sinh \eta)^{|n|-2}}{(1+\cosh \eta)^{|n|}} (d\eta^2+2i\frac{n}{|n|} \sinh \eta d\eta d\tau-\sinh^2\eta d\tau^2) \; , \quad |n|>1 \; 
\end{equation}
which are not correctly accounted for by the heat kernel and which must be treated separately.\footnote{The analysis in \cite{Sen:2012cj} proceeds by dimensionally reducing NHEK along the angular directions. The gauge field arising from the $\partial_\phi$ isometry is then argued to furnish another set of vector zero modes analogous to the AdS$_2$ perturbations \eqref{eq:vec}, although no explicit formula is presented. The naive guess corresponding to $\xi^{(n)} \propto \Phi_n(\tau,\eta)\partial_\phi$ with $\Phi_n$ defined in \eqref{eq:PhiN} does not satisfy the harmonic gauge condition and is not a zero mode of the operator \eqref{eq:NhekKinOp}. Note that $\xi^{(n)} \propto \Phi_n(\tau,\eta)\partial_\phi$ already appears as part of the diffeomorphism \eqref{diff} that generates the tensor zero modes \eqref{eq:ZmodesKERR}. A similar twist of conformal transformations by large $U(1)$ gauge transformations in the AdS$_2$ throat is described in \cite{Hartman:2008dq}. Regardless, as argued in \cite{Iliesiu:2022onk}, these vector zero modes are only expected to contribute in a specific choice of ensemble, and so are irrelevant to the universal, ensemble-independent $\frac32 \log T$ correction that we compute in this paper.}
These metric perturbations are zero modes precisely because they are generated by large diffeomorphisms left unfixed by harmonic gauge \eqref{eq:GF}. In other words, they obey $h^{(n)}\propto\mathcal{L}_{\xi^{(n)}}g_{\text{NHEK}}$,
with the vector field given by
\begin{equation} \label{diff}
    \xi^{(n)} = e^{in \tau} \tanh^{|n|}(\eta/2) \left( -\frac{|n|(|n|+\cosh \eta) + \sinh^2\eta}{\sinh^2 \eta} \partial_\tau + \frac{in(|n|+\cosh\eta)}{\sinh \eta} \partial_\eta +\frac{i(\cosh \eta+1+|n|-n^2)}{\cosh\eta+1} \partial_\phi \right) \; 
\end{equation}
and satisfying $\Box \xi^{(n)} = 0$. Repackaging these modes $\xi = \sum_n f_n \xi^{(n)}$ and defining $f(\tau) = \sum_n f_n e^{in\tau}$, one finds the large $\eta$ behavior
\begin{equation}\label{eq:DiffS1}
    \xi \approx -f(\tau) \partial_\tau +f'(\tau)\partial_\eta +if(\tau)\partial_\phi \; .
\end{equation}
These diffeomorphisms therefore correspond to boundary time reparametrizations that send $\tau \to \tau - f(\tau)$, $\eta \to \eta + f'(\tau)$, and $\phi \to \phi +if(\tau)$ and  resemble vector fields appearing in Kerr/CFT \cite{Guica:2008mu,Castro:2009jf,Bredberg:2011hp,Hartman:2008dq}.
The path integral \eqref{eq:1loop} is therefore proportional to an integral over the (infinite-dimensional, non-compact) coset Diff$(S^1)/SL(2,\mathbb{R})$. The quotient by 
$SL(2,\mathbb{R})$ arises because the $n=0,\pm 1$ perturbations, which would correspond to diffeomorphisms generated by \eqref{eq:NHEKSl2}
(i.e. $\mathcal{L}_{L_{\pm 1},L_0}\bar{g}$)
vanish due to the isometries of the background metric. This symmetry breaking pattern, explicated in \cite{Maldacena:2016upp}, is known to control many aspects of the near-extremal thermodynamics of spherically symmetric black holes. Since the mode \eqref{eq:DiffS1} costs no action and has infinite volume, the one-loop approximation to the path integral therefore suffers from an infrared divergence
\begin{equation}\label{eq:NhekIRdiv}
 Z\propto  \int \displaylimits_{\text{Diff}(S^1)/SL(2,\mathbb{R})}[Df(\tau)] \; =\infty \; 
\end{equation}
which is totally independent of any UV completion and completely controlled by the low energy fields in the model.  The (ultralocal) measure $[Dh]$ induces a measure on Diff$(S^1)/SL(2,\mathbb{R})$ whose dependence on $S_0$ can be determined exactly. This in turn determines the log $S_0$ correction to the entropy coming from the zero modes as reported in \cite{Sen:2012cj}. The NHEK path integral  itself is however infrared divergent and ill-defined.

\subsection{The not-NHEK Metric}

The infinity \eqref{eq:NhekIRdiv} is an infrared divergence, which arises from low energy modes of low energy fields, and is therefore a physical effect. Its existence calls into question the basic assumption that the NHEK path integral computes the zero-temperature black hole partition function. One way to settle this question would be to define and compute the finite temperature partition function for the black hole and then to take the $T\to 0$ limit. In other words, we would like to know if $Z_{\text{NHEK}}=\lim_{T\to 0}Z[T]_{\text{Black Hole}}$ when quantum fluctuations are taken into account.

The issue is that $``Z[T]_{\text{Black Hole}}$'' is itself difficult to define, let alone compute, at finite temperature in asymptotically flat space. The most obvious definition would involve a Euclidean path integral with the standard asymptotically flat boundary conditions and a periodic identification of asymptotic Euclidean time. This computation involves integrating over fluctuations far from the black hole, whose contributions have to be disentangled from the trace over the black hole microstates. The calculation cannot be performed explicitly for the Kerr black hole beyond the leading saddle-point approximation, 
although Sen was able to derive predictions for the logarithmic corrections to the entropy of nonextremal black holes in \cite{Sen:2012dw,Bhattacharyya:2012wz}  through a careful consideration of the heat kernel and zero modes. To extract these corrections, Sen utilized a scaling limit where the temperature grows at the same rate as the angular momentum, so taking the $T \to 0$ limit of his computation at fixed $J$ is ill-defined and there is no way to compare with the NHEK partition function.\footnote{In \cite{H:2023qko,Anupam:2023yns,Iliesiu:2021are}, the authors compute an index (which is temperature independent) using the full asymptotic geometry and find agreement with the results of the near-horizon analysis. In these models supersymmetry actually regularizes the divergence \eqref{eq:NhekIRdiv} so there is no puzzle to begin with. }

The authors of \cite{Iliesiu:2022onk,Banerjee:2023quv} adopt a different definition of $Z[T]_{\text{Black Hole}}$ for the low temperature Reissner-Nordstr\"{o}m black hole, and their main conclusion is that in the absence of supersymmetry, $\lim_{T\to 0}Z[T]_{\text{Black Hole}} \neq Z_{\text{AdS$_2$}\times \text{S$^2$}}$. We review these calculations in Appendix \ref{app:AdS2} and extend the analysis to the Kerr black hole below. 

The main assumption underlying their definition of $Z[T]_{\text{Black Hole}}$ is that, for small temperatures, one can simply correct the throat geometry \eqref{eq:EucNHEK} rather than perform the full asymptotically flat path integral. It is not obvious that this is mathematically equivalent to taking the small $T$ limit of the full path integral with asymptotically flat boundary conditions, but it seems physically plausible that the leading corrections to the low temperature thermodynamics arise from dynamics near the throat. Either way, since the full finite temperature black hole certainly does not support the infinite set of zero modes \eqref{eq:ZmodesKERR}, it is clear that the IR divergence will disappear in either prescription. Whether the form of the correction is the same is less obvious.

At a technical level, the prescription amounts to performing the diffeomorphism \eqref{eq:throat diff}, and then expanding the resulting metric in powers of $T$ instead of taking the strict $T\to 0$ limit. The leading term is of course the NHEK metric \eqref{eq:EucNHEK}. The subleading term represents a (non-normalizable) gravitational perturbation of NHEK whose nonlinear completion is the asymptotically flat finite temperature Kerr black hole. This $O(T)$ correction to the Wick-rotated metric (denoted by $\delta g$ in \eqref{eq:IntroNotNHEK}) is given by
\begin{equation} \label{eq:1stCORR}
\begin{aligned}
    \frac{\delta g_{\mu \nu}dx^\mu dx^\nu}{4 \pi J^{3/2} T} =& (1+\cos^2 \theta) (2+\cosh \eta) \tanh^2 \frac\eta2 (d\eta^2 - \sinh^2 \eta d\tau^2) + \sin^2 \theta \cosh \eta (d\eta^2 + \sinh^2 \eta d\tau^2)+2 \cosh \eta \,d\theta^2 \\
    &+2 \frac{\sin^2 \theta}{1+\cos^2 \theta} (\cosh \eta - 1) \left( (\sin^2 \theta \sinh^2 \eta - 3)  - 4 \frac{\cos^2 \theta}{1+\cos^2 \theta} \cosh \eta (\cosh \eta -1) \right) d\tau^2 \\
    &+2i \frac{\sin^2 \theta}{1+\cos^2 \theta}\left((\sin^2 \theta \sinh^2 \eta - 3) - 8 \frac{\cos^2 \theta}{1+\cos^2\theta} \cosh \eta (\cosh \eta - 1) \right) d\tau d\phi \\
    &+8 \cosh \eta \frac{\sin^2 \theta \cos^2 \theta}{(1+\cos^2\theta)^2} d\phi^2  \; .
\end{aligned}
\end{equation}
Note that the last three lines in the correction actually correspond to the $O(T)$ terms in $g_{\phi \phi} (d \phi + a_\tau d\tau)^2$ with
\begin{equation}
\begin{aligned}
    & a_\tau =\bar{a}_\tau + \delta a_\tau= -i (\cosh \eta -1) + i \pi J^{1/2} T (\sin^2 \theta \sinh^2 \eta - 3)\; , \\
    & g_{\phi \phi} =\bar{g}_{\phi \phi} + \delta g_{\phi \phi}= 4 J \frac{\sin^2 \theta}{1+\cos^2 \theta} \left(1 + 8 \pi J^{1/2}T \cosh \eta \frac{\cos^2 \theta}{1+\cos^2 \theta}  \right) \; .
\end{aligned}
\end{equation}
Importantly, $\mathcal{L}_{L_0} \delta g  =0$ while $\mathcal{L}_{L_{\pm}} \delta g = \textbf{g}_{\pm}$ with $\textbf{g}_{\pm}$ non-normalizable, so that one is still not integrating over the $n=0,\pm 1$ modes that would correspond to $SL(2,\mathbb{R})$ diffeomorphisms. As noted in \eqref{eq:zmnotNHEK}, the NHEK zero modes  \eqref{eq:ZmodesKERR} do not result from diffeomorphisms of this corrected geometry, so we expect their eigenvalues to pick up corrections of order $T$.

\subsection{Eigenvalue Corrections to the Extremal Zero Modes and log$\,T$ Corrections to the Entropy}
The correction \eqref{eq:1stCORR} to the NHEK metric induces a correction $\delta D$ to the NHEK kinetic operator $\bar{D}$ in \eqref{eq:NhekKinOp}. This in turn modifies the extremal eigenfunctions $h^0$ and their eigenvalues $\Lambda^0$. Expanding everything to first order in $T$
\begin{equation}
    (\bar{D} + \delta D)(h^0_{n} + \delta h_{n}) = (\Lambda_{n}^0 + \delta \Lambda_{n}) (h^0_{n} + \delta h_{n})
\end{equation}
and isolating the $O(T)$ terms, we get
\begin{equation}
    \bar{D} \delta h_n + \delta D h_n^0 = \Lambda _n ^0 \delta h_n + \delta \Lambda_n h_n^0 \; .
\end{equation}
Taking the inner product with $h_m^0$, using orthonormality of the $0^{th}$ order eigenfunctions, and restoring indices, the $1^{st}$ order correction to the eigenvalue takes the form
\begin{equation}\label{eq:correction}
    \delta \Lambda_n = \int d^4 x \sqrt{\bar{g}} (h_n^0)_{\alpha \beta} \delta D^{\alpha \beta , \mu \nu} (h_n^0)_{\mu \nu} \; .
\end{equation}
The corrected one loop determinant is therefore
\begin{equation}
    \log Z = - \frac12 \sum_n \log (\Lambda_n^0 + \delta \Lambda_n) \;,
\end{equation}
with $\delta \Lambda_n \sim T$. This makes it clear that modes which have non-zero extremal eigenvalues ($\Lambda_n^0 \neq 0$) produce subleading, polynomially-suppressed temperature dependence relative to the modes whose extremal eigenvalues vanish.
The latter are precisely the real and imaginary parts of \eqref{eq:ZmodesKERR}.
The leading order correction to the kinetic operator is
\begin{equation}
    \delta D^{\alpha \beta, \mu \nu}=-\frac{1}{16\pi}\delta\left(\frac14 g^{\alpha \mu} g^{\beta \nu} \Box - \frac18 g^{ \mu \nu} g^{\alpha \beta } \Box +\frac12 R^{\alpha \mu \beta \nu}   \right) 
\end{equation}
with  $g_{\text{not-NHEK}} = \bar{g} + \delta g$. The operator itself is utterly intractable, but the quantity \eqref{eq:correction} with $h_n^0$ given by the real and imaginary parts of \eqref{eq:ZmodesKERR} simplifies dramatically and takes the form
\begin{equation} \label{corr_e}
\delta \Lambda_n = \frac{3 n T}{64J^{1/2}} \; , \qquad n\geq2 \; .
\end{equation}
As an aid to readers we record the intermediate result
\begin{align}
   & \int d^4 x \sqrt{\bar{g}} (h_n^0)_{\alpha \beta} \delta D^{\alpha \beta , \mu \nu} (h_n^0)_{\mu \nu}=-\frac{3n(n^2-1) T}{128 J^{1/2}}\int_0^\infty d\eta\Big[  16(\pi-2)   \coth \eta  \, \text{csch}^2\eta \, \tanh ^{2 n}\left(\frac{\eta }{2}\right)
    -\\&  \text{csch}^3\eta \,  \text{sech}^4\left(\frac{\eta }{2}\right) \left((\pi -2) \cosh 3 \eta +(4 (n-2) n+7 \pi -30) \cosh \eta -2 (n-2 \pi +4) \cosh 2 \eta +2 n (4 n+7)+4 \pi \right) \tanh ^{2 n}\left(\frac{\eta }{2}\right)
    \Big]\; .\notag
\end{align}
In this expression the first term is the Riemann contribution and the second term comes from the Laplacian.

The contribution of the extremal zero modes to the not-NHEK partition function is therefore 
\begin{equation} \label{corr_eigen_J}
\begin{aligned}
    \delta \log Z &= 2 \cdot (-1/2)\sum_{n\geq 2} \log \delta \Lambda_n 
    = \log \left( \prod_{n \geq 2}  \frac{64J^{1/2}}{3 n T} \right) \; ,
\end{aligned}
\end{equation}
where the factor of $2$ comes from including the identical contributions from the real and imaginary parts of the perturbations.  Using zeta function regularization to compute the infinite product
\begin{equation}
    \prod_{n \geq 2} \frac{\alpha}{n} = \frac{1}{\sqrt{2\pi}} \frac{1}{\alpha^{3/2}} \,
\end{equation}
the final answer takes the form 
\begin{equation} \label{corr_Z}
\delta \log Z= \log \left( \frac{\sqrt{27}}{512 \sqrt{2\pi}} \frac{T^{3/2}}{J^{3/4}} \right) 
    \sim \frac32 \log T \; .
\end{equation}
We conclude that at low temperatures
\begin{equation}
    Z[T]_{\text{Black Hole}}\sim T^{3/2}e^{S_0} + \textit{higher order terms} \; .
\end{equation}
It remains to understand the regime of validity of this expression and its physical content. Obviously, once the small temperature dependent prefactor begins competing with the large temperature independent exponential, the approximation is not valid. This occurs when $T^{3/2}\sim e^{-S_0}$. Below this temperature, the partition function is so small that other saddles will begin competing with the computation performed here. Similarly, when $T\sim J^{-1/2}$ the linear term in \eqref{eq:LinT} competes with the leading $S_0$ term and the near-extremal approximation breaks down. Equivalently, the correction term \eqref{eq:1stCORR} becomes as large as the NHEK metric \eqref{eq:NHEK_sen} throughout the throat and the small-$T$ approximation of the geometry completely breaks down.  

The fact that the partition function vanishes as $T\to 0$ means that $\rho(E)\to 0$ as $E\to 0$. There is no exponential ground state degeneracy or thermodynamic mass gap \cite{Preskill:1991tb}. Rather, the would-be ground states are spread out over a dense energy band above the vacuum. Hence, standard thermodynamics still applies in the range ${J^{a_1} e^{-a_2 S_0}\lesssim T\lesssim J^{-1/2}}$, where $a_1,a_2$ are expected to be $O(1)$ numbers.

\subsection*{Acknowledgements}
We thank A. Castro, L. Iliesiu and A. Sen for useful discussions. This work  is supported by DOE grant de-sc/0007870 and  the John Templeton Foundation via the Harvard Black Hole Initiative    
and the  Marie Sk\l odowska-Curie Global Fellowship (ERC Horizon 2020 Program) SPINBHMICRO-101024314 for CT. CT also  acknowledges support from the Simons Center for Geometry and Physics, Stony Brook University at which some of the research for this paper was performed.

\appendix

\section{AdS$_2\times S^2$ Calculations}\label{app:AdS2}
In this appendix we review the computations of \cite{Iliesiu:2022onk,Banerjee:2023quv} for the extremal Reissner-Nordstr\"om black hole. The Einstein-Maxwell action with Gibbons-Hawking and electromagnetic boundary terms is
\begin{equation}
    I=-\frac{1}{16\pi} \int d^4x \sqrt{g}(R-F_{\mu \nu} F^{\mu \nu}) + \frac{1}{8\pi}\int d^3x \sqrt{\gamma}(K-2n_\mu A_\nu F^{\mu \nu}) \; ,
\end{equation}
and the Reissner-Nordstr\"om metric takes the form
\begin{equation}
    ds^2=-\left(1-\frac{2M}{\hat{r}} + \frac{Q^2}{\hat{r}^2} \right)d\hat{t}^2 + \left(1-\frac{2M}{\hat{r}} + \frac{Q^2}{\hat{r}^2}\right)^{-1}d\hat{r}^2 + \hat{r}^2(d\theta^2+\sin^2\theta d\phi^2) \; .
\end{equation}
The inner and outer horizons are located at the radial coordinates
\begin{equation}\label{rm_rel}
    r_{\pm}=M\pm \sqrt{M^2-Q^2} \; , \qquad  r_-r_+ = Q^2\; .
\end{equation}
The background electric field and Hawking temperature are given by
\begin{equation}
    A=Q\left(\frac{1}{r_+}-\frac{1}{\hat{r}}\right)d\hat{t} \; , \qquad F=\frac{Q}{\hat{r}^2} d\hat{r} \wedge d\hat{t}  \;,  \qquad     T = \frac{r_+^2-Q^2}{4\pi r_+^3} \; .
\end{equation}
At small temperatures above extremality,
\begin{eqnarray}
M(T,Q) &= &    Q+2 \pi ^2 Q^3 T^2+16 \pi^3 Q^4 T^3 + 126 \pi^4 Q^5 T^4+\dots \;  \\
r_+(T,Q) &= & Q+2 \pi  Q^2 T+ 10 \pi ^2 Q^3 T^2+ 64 \pi^3 Q^4 T^3+  462 \pi ^4 Q^5 T^4 \dots \; .
\end{eqnarray}
If we take the scaling limit
\begin{align}\label{eq:RNscale}
	\hat{t} = \frac{t}{2\pi T} \; , \qquad
	\hat{r} = r_+(T) + 2\pi Q^2T(\cosh\eta-1) \; , \qquad T \to 0 \; 
\end{align}
with $Q$ fixed, then the metric becomes
\begin{equation}
    ds^2=Q^2 (- \sinh^2 \eta\, \text{d}t^2+ \text{d}\eta^2)  +Q^2(d\theta^2+\sin^2\theta d\phi^2) \; ,
\end{equation}
which is $AdS_2 \times S^2$ with a constant background electric field
\begin{equation}
    A=Q(\cosh \eta-1)dt \;, \qquad \, F=Q\sinh \eta \,d\eta \wedge dt \; .
\end{equation}
The linearized kinetic term on this background is \cite{Sen:2012kpz,Bhattacharyya:2012wz}
\begin{align}\label{eq:RNkin}
-16 \pi h_{\alpha \beta}D^{\alpha \beta ,\mu \nu}h_{\mu \nu} &= h_{\alpha \beta}\left(\frac14 \bar{g}^{\alpha \mu } \bar{g}^{\beta \nu} \bar{\Box} -\frac18 \bar{g}^{\alpha \beta } \bar{g}^{\mu\nu} \bar{\Box} +\frac12 \bar{R}^{\alpha \mu \beta \nu} +\frac12 \bar{R}^{\alpha \mu} \bar{g}^{\beta \nu} - \frac12 \bar{R}^{\alpha \beta } \bar{g}^{\mu\nu}  \right) h_{\mu \nu}\\
&+ h_{\alpha \beta}\left(\frac18 \bar{F}^2\left(2\bar{g}^{\alpha \mu} \bar{g}^{\beta \nu}-\bar{g}^{\alpha \beta } \bar{g}^{\mu \nu}\right)-\bar{F}^{\alpha \mu} \bar{F}^{\beta \nu}-2 \bar{F}^{\alpha \gamma} \bar{F}^\mu_{\;\;\;\gamma} \bar{g}^{\beta \nu}+\bar{F}^{\alpha \gamma} \bar{F}^\beta_{\;\;\;\gamma} \bar{g}^{\mu \nu}\right)h_{\mu \nu} \; \nonumber .
\end{align}
$D$ has a number of (tensor and vector) zero modes which we treat separately below.

\textbf{Tensor modes}. After Wick rotation $t=-i\tau$, the operator defined in \eqref{eq:RNkin} admits tensor zero modes of the form \cite{Camporesi:1994ga}
\begin{equation}\label{eq:zmRN}
     h_{\mu \nu}^{(n)}dx^\mu dx^\nu=  \frac{\sqrt{|n|(n^2-1)}}{2\pi}\frac{(\sinh \eta)^{|n|-2}}{(1+\cosh \eta)^{|n|}}e^{in \tau} (\text{d}\eta^2+2i\frac{n}{|n|}\sinh \eta \,\text{d}\eta \text{d}\tau-\sinh^2\eta  \text{d}\tau^2) \; .
\end{equation}

The one-loop approximation to the partition function is therefore infrared divergent. To regulate this divergence, the authors of \cite{Iliesiu:2022onk,Banerjee:2023quv} keep the subleading term in the expansion \eqref{eq:RNscale}. The corrected metric has the form 
\begin{equation}
 \frac{\delta g_{\mu \nu}dx^\mu dx^\nu}{4\pi Q^3T}=   (2+\cosh \eta)\tanh^2\frac{\eta}{2}(d\eta^2-\sinh^2\eta d\tau^2) +  \cosh \eta (d\theta^2+ \sin^2\theta d\phi^2)
\end{equation}
while the correction to the field strength is
\begin{equation}
    \delta A = 2 \pi i Q^2 T \sinh^2 \eta d\tau \;, \qquad \, \delta F = 4\pi i Q^2T \sinh \eta \cosh \eta \, d\eta \wedge d\tau \; .
\end{equation}
Simple power-counting indicates that $\delta D^{\alpha \beta;\mu \nu}\sim T/Q^5$, so that the action for these modes scales like $T/Q$. Calculating the coefficient carefully, one finds
\begin{equation}
 \delta \Lambda_n=  \int \sqrt{\bar{g}}d^4x\, (h^0_{n})_{\alpha \beta} \left(\delta D^{\alpha \beta;\mu \nu}\right)(h^0_{n})_{\mu \nu} =\frac{n T}{16Q} \; , \qquad n\geq 2 \; .
\end{equation}
Including a factor of $2$ coming from the real and imaginary parts of the perturbation, and using zeta function regularization, one finds the following contribution:
\begin{equation} \label{contrib_tens}
\delta \log Z = 2 \cdot (-1/2)\sum_{n\geq 2} \log \delta \Lambda_n 
     =  \log \left( \prod_{n \geq 2}  \frac{16 Q}{ n T} \right) =  \log \left( \frac{1}{ 64 \sqrt{2\pi}} \frac{T^{3/2}}{Q^{3/2}} \right)  \; . 
\end{equation}

\textbf{Vector modes.} There are additional contributions from the vector zero modes of the metric given by \cite{Iliesiu:2022onk,Banerjee:2023quv}
\begin{equation}\label{eq:vec}
h_{i \mu} = \frac{1}{\sqrt{2}}\epsilon_{ij} \partial^{j} Y_{l}^m (\theta) v_{\mu} \; .
\end{equation}
Here $Y_{l}^m$ are the spherical harmonics, $i,j$ denote the  coordinates $\theta, \phi$,  $\epsilon_{ij}$ is the 2d Levi Civita symbol and 
\begin{equation}\label{eq:PhiN}
v_{\mu} = \partial_{\mu} \Phi_n \qquad \Phi_n = \frac{1}{\sqrt{2 \pi |n|}} \left( \frac{\sinh \eta}{1 + \cosh \eta}\right)^{|n|} e^{i n \tau} \qquad n = \pm1, \pm 2 ...
\end{equation}
The eigenvalue correction before integration over $\eta$ assumes the following form, for each mode
\begin{equation}
  \delta \Lambda_n= \frac{T  n (n+1)}{32 Q}\int_{0}^{\infty} d\eta \left[  \text{csch}\left(\frac{\eta}{2}\right) \, \text{sech}^5\left(\frac{\eta}{2}\right)\tanh ^{2 n}\left(\frac{\eta}{2}\right) \left((n-1) \cosh \eta+2 n+1\right)  \right]\,,
\end{equation}
which, upon integration, yields
\begin{equation}
    \delta \Lambda_n= \frac{n T}{16 Q}\, \qquad n \geq 1.
\end{equation}
Including a factor of $2$ coming from the real and imaginary parts of the perturbation, and using zeta function regularization, one finds
\begin{equation}
\begin{aligned} \label{contrib_vector}
    \delta \log Z_v &= 6 \cdot (-1/2)\sum_{n\geq 1} \log \delta \Lambda_n 
     = 3 \log \left( \prod_{n \geq 1}  \frac{16 Q}{  n T} \right) = 3 \log \left( \frac{1}{\sqrt{32 \pi}} \frac{T^{1/2}}{Q^{1/2}} \right) \; .
\end{aligned}
\end{equation}
The combined contribution of vector and tensor modes amounts to 
\begin{equation}
   \delta Z = \delta Z_t + \delta Z_v =  \log \left( \frac{1}{ 64 \sqrt{2\pi}} \frac{T^{3/2}}{Q^{3/2}} \right) +3  \log \left( \frac{1}{\sqrt{32 \pi}} \frac{T^{1/2}}{Q^{1/2}}\right) \; .
\end{equation}
There is also an additional contribution from the vector zero modes of the gauge field. However, as described in \cite{Iliesiu:2022onk}, different subsets of the zero modes contribute depending on which ensemble is being considered, but the tensor zero modes of the metric always contribute the universal factor $\frac32 \log T$.

\bibliography{Bib}
\bibliographystyle{utphys}
\end{document}